\documentclass[fleqn,usenatbib]{mnras}

\usepackage{newtxtext,newtxmath}

\usepackage[T1]{fontenc}
\usepackage{ae,aecompl}
\usepackage[utf8]{inputenc}
\usepackage{bm}

\usepackage{graphicx}	
\usepackage{amsmath}	
\usepackage[acronym,shortcuts]{glossaries} 
\usepackage{booktabs}
\glsdisablehyper 
\usepackage{ulem} 
\usepackage{xspace}
\usepackage[dvipsnames]{xcolor} 
\usepackage{subfig}

\newcommand{\Rsun}{\ensuremath{\,\rm{R}_{\odot}}\xspace}

\newcommand{\kms}{\ensuremath{\,\rm{km}\,\rm{s}^{-1}}\xspace}
\newcommand{\Msun}{\ensuremath{\,\rm{M}_{\odot}}\xspace}

\newcommand{\yr}{\ensuremath{\,\mathrm{yr}}\xspace}

\newcommand{\erg}{\ensuremath{\,\mathrm{erg}}\xspace}
\newcommand{\Phantom}{{\scshape phantom}\xspace }
\newcommand{\SPLASH}{{\scshape splash}\xspace }
\newcommand{\MESA}{{\scshape mesa}\xspace }

\newcommand{\molH}{H${}_2$\xspace }
\newcommand{\irecA}{``He + H + H${}_2$''\xspace}
\newcommand{\irecB}{``He + H''\xspace}
\newcommand{\irecC}{``He''\xspace}
\newcommand{\irecD}{``None''\xspace}
\newcommand{\irecE}{``None, fixed $\mu$''\xspace}

\newacronym{RSG}{RSG}{red supergiant}

\newacronym{CE}{CE}{common-envelope}

\newacronym{SPH}{SPH}{smoothed particle hydrodynamics}
\newacronym[longplural={equations of state}, plural={EoSs}]{EoS}{EoS}{equation of state}
\newacronym{LTE}{LTE}{local thermodynamic equilibrium}

\title[Common envelopes II: The distinct roles of H and He recombination]{Common envelopes in massive stars II: The distinct roles of hydrogen and helium recombination}

\author[M. Y. M. Lau et al.]{Mike Y. M. Lau,$^{1,2,3}$\thanks{E-mail: mike.lau@monash.edu (MYML)},
	    Ryosuke Hirai$^{1,2}$,
		Daniel J. Price$^{1}$, 
		Ilya Mandel$^{1,2}$
	\\
	$^1$ School of Physics and Astronomy, Monash University, Clayton, Victoria 3800, Australia\\
	$^2$ OzGrav: The ARC Centre of Excellence for Gravitational Wave Discovery, Australia\\
	$^3$ Center for Computational Astrophysics, Flatiron Institute, 162 Fifth Avenue, New York, NY 10010, USA\\
}

\date{Accepted XXX. Received YYY; in original form ZZZ}

\pubyear{2022}

\begin{document}
\label{firstpage}
\pagerange{\pageref{firstpage}--\pageref{lastpage}}
\maketitle

\begin{abstract}
	The role of recombination during a common-envelope event has been long debated. Many studies have argued that much of hydrogen recombination energy, which is radiated in relatively cool and optically-thin layers, might not thermalise in the envelope. On the other hand, helium recombination contains $\approx 30$ per cent of the total recombination energy, and occurs much deeper in the stellar envelope. We investigate the distinct roles played by hydrogen and helium recombination in a common-envelope interaction experienced by a 12\Msun red supergiant donor. We perform adiabatic, 3D hydrodynamical simulations that (i) include hydrogen, helium, and H${}_2$ recombination, (ii) include hydrogen and helium recombination, (iii) include only helium recombination, and (iv) do not include recombination energy. By comparing these simulations, we find that the addition of helium recombination energy alone ejects 30 per cent more envelope mass, and leads to a 16 per cent larger post-plunge-in separation. Under the adiabatic assumption, adding hydrogen recombination energy increases the amount of ejected mass by a further 40 per cent, possibly unbinding the entire envelope, but does not affect the post-plunge separation. Most of the ejecta becomes unbound at relatively high ($>70$ per cent) degrees of hydrogen ionisation, where the hydrogen recombination energy is likely to expand the envelope instead of being radiated away.
\end{abstract}



\section{Introduction}
The ability for recombination energy to help eject expanding stellar envelopes was initially explored in the context of planetary nebula formation from single stars \citep{Lucy67,Roxburgh67,Paczsynski+Ziolkowski68,Han+94,Harpaz98}. Recombination has since been studied in \ac{CE} evolution \citep{Han+95}, where the envelope of an evolved giant star expands and cools after interacting with an engulfed stellar companion. Sufficiently cooled layers of the envelope recombine, releasing energy that may further inflate the \ac{CE} and aid in its ejection.

There have been observational suggestions that recombination energy may help eject \acp{CE}, based on reconstructing the \ac{CE} parameters of wide post-\ac{CE} binaries \citep{Webbink08,Zorotovic+10,Davis+12,Rebassa-Mansergas+12,Iaconi&DeMarco19}. Some population synthesis studies also find that models with large \ac{CE} efficiency parameters (``$\alpha_\mathrm{CE}$''), which imply additional energy sources beyond orbital energy, such as recombination, are required to match the period distributions of, e.g., subdwarf B (sdB) stars \citep{Han+03} and double white dwarfs \citep{Nelemans+00}. The role of recombination energy in \ac{CE} evolution has been studied using 3D hydrodynamical simulations, 1D simulations, and analytical arguments based on (often) static, 1D stellar models.

Hydrodynamical simulations typically explore the additional amount of envelope mass that may be ejected when recombination energy is included and assumed to fully thermalise in the envelope. The latter assumption arises from the adiabatic nature of these simulations; simulating \ac{CE} evolution in 3D with radiation hydrodynamics remains computationally challenging \cite[but, see][]{Ricker+2018}. Whereas standard 3D \ac{CE} simulations that do not model recombination typically result in ejecting few tens of percent of the envelope \citep[but, see][who focused on the last one percent of the dynamical plunge]{Law-Smith+20}, including recombination energy has resulted in possible complete ejection \citep[e.g.][]{Nandez+15,Ivanova&Nandez16,Reichardt+20,Sand+20,Lau+22,Gonzalez-Bolivar+22}. Some of these studies also found that most of the helium recombination energy may be used to expand the envelope, whereas a significant fraction of hydrogen recombination energy is released in material that has already become unbound. The effect of recombination energy on the final orbital separation is less clear. Both \cite{Ivanova&Nandez16} and \cite{Reichardt+20} compared \ac{CE} simulations that include recombination energy with simulations that exclude it, reporting little difference in the final orbital separation. However, \cite{Sand+20} and \cite{Gonzalez-Bolivar+22} both found a larger final separation when including recombination energy in \acp{CE} with asymptotic giant branch donors, despite these simulations also ejecting a larger fraction of the envelope mass. This was also reported by \cite{Lau+22} for a 12 \Msun \ac{RSG} donor, where including recombination energy resulted in a 20 per cent larger final separation.

However, interpreting these results requires caution, as the adiabatic assumption breaks down in the optically thin outer layers of the hydrogen partial ionisation zone, where gas is allowed to cool radiatively. Hydrogen recombination radiation that is released in these regions may diffuse or stream away instead of driving expansion. More generally, including energy transport becomes important when simulating \ac{CE} evolution past the dynamical spiral-in, and is essential for the self-regulated phase \citep{Meyer+Meyer-Hofmeister79,Ivanova+15,Clayton:2017}. In 1D hydrodynamical simulations, radiation transport may be included, and convective energy transport may be treated using mixing length theory \citep[e.g.,][who use the dynamical module of \MESA]{Fragos+2019}. However, caution is needed in discerning 1D artefacts, such as the effect of instantaneous energy homogenisation in radial shells.

There has also been a number of analytical investigations based on static 1D stellar models that scrutinise the ability for recombination energy to aid in envelope ejection. One approach is based on comparing the expansion and energy transport time-scales in the envelope. With this approach, \cite{Sabach+17} estimated that convection should carry out half of the helium recombination energy to the photosphere, where it can be radiated away. \cite{Grichener+18} performed a similar analysis for an expanding envelope, claiming that most recombination energy is radiated away. \cite{Ivanova18} performed an analysis based on comparing the radiative, convective, and recombination energy fluxes, arguing instead that neither radiative diffusion nor convection may efficiently transport away hydrogen recombination energy released in bound layers of the envelope. Recently, \cite{Wilson&Nordhaus22} argue that convection is not efficient in carrying out the orbital energy injected by the companion in massive star \acp{CE}.

The combined efforts of 3D simulations, 1D simulations, and analytical arguments broadly converge onto several conclusions. Firstly, recombination energy can be a significant energy source for envelope ejection if it is thermalised, although this can depend on the chosen boundary between the ejected envelope and the remaining bound mass, the so-called ``bifurcation point''. Secondly, while helium recombination energy is injected in optically-thick layers, the ability for hydrogen recombination energy to thermalise in the envelope is less certain. Despite this, there has not yet been a hydrodynamical simulation that studies the isolated role of helium recombination in the absence of hydrogen recombination, which would serve as a useful lower limit to the role of recombination energy.

In this work, we perform and compare simulations that include different recombination energy sources: (i) Helium and hydrogen recombination (including molecular recombination to \molH), (ii) Helium and hydrogen atomic recombination only, (iii) Helium recombination only, and (iv) No recombination. Since our simulations are adiabatic, they implicitly assume recombination radiation is locally thermalised. By comparing (ii) and (iii), we disentangle the effects of hydrogen and helium recombination in a \ac{CE}. This is difficult in past studies, which have only compared simulations including full recombination and without any recombination (i.e., comparing (i) and (iv) in our designation). In those simulations, it is unclear whether a partially-ionised layer that becomes unbound would have eventually been ejected through hydrodynamical interactions even in the absence of hydrogen recombination energy.

This paper is organised as follows. We summarise our setup in Section \ref{sec:methods} and explain our treatment of recombination in Section \ref{subsec:recombination}. In Section \ref{sec:results}, we compare the fraction of unbound envelope mass (\ref{subsec:unbound}) and final separation (\ref{subsec:sep}) between our simulations. In Section \ref{sec:discussion}, we discuss the effect of recombination on the ejecta structure (\ref{subsec:ejecta}) and the ability for recombination energy to thermalise in the envelope (\ref{subsec:transport}). We summarise our findings in Section \ref{sec:conclusion}.

\section{Methods} \label{sec:methods}
Our simulations have the same setup as our previous work (Paper I), \cite{Lau+22}, except for a new implementation of recombination physics (see Section \ref{subsec:recombination}). In this Section, we therefore summarise our setup, and refer the reader to Section 2 of \cite{Lau+22} for details.

We simulate a \ac{CE} experienced by a 12\Msun \ac{RSG} donor with a 3\Msun companion, using the same donor density profile as in \cite{Lau+22}. We use a newer version of the \ac{SPH} code \Phantom \citep[v2022.0.1,][]{Price+18}, where we have introduced the ability to simulate fluids with non-uniform composition and a new \ac{EoS} prescription that allows different recombination energy sources to be included separately (see Section \ref{subsec:recombination}). We resolve the donor star with $2\times 10^6$ \ac{SPH} particles, which was our default in \cite{Lau+22}. We check convergence by comparison with a set of simulations carried out at lower resolution (using $2\times10^5$ \ac{SPH} particles).

We set up an initially circular orbit with separation 988\Rsun, such that the donor's radius (618\Rsun) exceeds its Roche radius by 30 per cent. The donor star is initially non-rotating, and we do not account for the effects of tidal distortion and thermal time-scale mass transfer, which are likely to modify the donor stellar profile during the onset of Roche-lobe overflow. \cite{Lau+22} demonstrated that the donor star maintains its original density structure for at least 90 times the surface free-fall time when simulated in the absence of a companion.

We perform our simulations on the 64-core AMD EPYC 7742 Rome processors on the Flatiron Institute Rusty computing cluster, and on the 64-core Intel Xeon Platinum 8358 Icelake processors on the Flatiron Popeye computing cluster. A single \ac{CE} simulation, using OpenMP parallelisation, consumes roughly 100 kcpu-hr, and requires 2--3 months of wall time. Our simulations conserve energy to within 0.04 per cent and angular momentum to within 0.02 per cent, due to our use of a single, global timestep for all particles.

\section{Treatment of recombination}
\label{subsec:recombination}
Atomic and molecular recombination are associated with an increase in mean molecular weight and, when in \ac{LTE}, an increase in thermal energy. These effects may be incorporated during \ac{EoS} evaluation in a hydrodynamics code. In particular, the ionisation potential manifests as an additional internal energy term. In our previous simulations that include recombination, we used the \MESA \ac{EoS} tables \citep{MESA1,MESA2,MESA3,MESA4,MESA5}, which are constructed from the OPAL \ac{EoS} \citep{Rogers+96,Rogers+Nayfonov02} and SCVH \ac{EoS} \citep{Saumon+95}. This tabulated \ac{EoS} includes the recombination energy available to various ionisation states. For a typical stellar composition, the most significant sources are ionised and neutral hydrogen (forming \molH) and singly- and doubly-ionised helium. In order to separately investigate the effects of hydrogen and helium recombination, which is the aim of this paper, we instead include the ionisation potential, $\varepsilon_\mathrm{ion}$, of each species analytically in the expression for specific internal energy, $u$:
\begin{align}
	u = \frac{3 k_\mathrm{B} T}{2\mu(x_i)m_\mathrm{u}} + \frac{a_\mathrm{rad}T^4}{\rho} + \varepsilon_\mathrm{ion}(x_i),
	\label{eq:eint}
\end{align}
where $\rho$ and $T$ denote the gas density and temperature, and $k_\mathrm{B}$, $a_\mathrm{rad}$, and $m_\mathrm{u}$ are the Boltzmann constant, radiation constant, and atomic mass unit, respectively. $\mu$ is the mean molecular weight, which is a function of chemical composition (determined by the hydrogen and helium mass fractions, $X$ and $Y$) and the ionisation/dissociation fractions $x_{1},...,x_{4}$ of $\mathrm{H_2}$, $\mathrm{H}$, $\mathrm{He}$, and $\mathrm{He^+}$, respectively. As the chemical composition is uniform in a \ac{RSG} convective envelope, we assume fixed hydrogen and helium mass fractions, $X=0.698$ and $Y=0.287$, respectively. 

The total specific recombination energy, or ionisation potential, is a sum of its contribution from different hydrogen and helium ionisation states,
\begin{align}
	\varepsilon_\mathrm{ion}(x_1,...,x_4) = 
	\varepsilon_\mathrm{H_2}x_1 + \varepsilon_\mathrm{H}x_2 + \varepsilon_\mathrm{He}x_3 + \varepsilon_\mathrm{He^+}x_4,
	\label{eq:epsilon}
\end{align}
where, from left to right, the terms on the RHS are the specific recombination energies available to H (forming \molH), H$^+$, He$^+$, and He$^{2+}$.

The ionisation potential of a given species depends on its ionisation fraction, which, in \ac{LTE}, depends on $\rho$ and $T$ via the Saha equations. Instead of solving Saha equations, we use accurate analytical fits of $x_i(\rho,T)$ to the \MESA \ac{EoS} table \citep{MESA1,MESA2,MESA3,MESA4,MESA5} to ensure fast computational evaluation. The analytical expressions and fitting coefficients are provided in Appendix C of \cite{Hirai+2020}. Each \ac{EoS} evaluation involves solving for $T$ in equation (\ref{eq:eint}). This poses a challenge for traditional root finders like the Newton-Raphson method, which tends to diverge near the inflection points in $\varepsilon_\mathrm{ion}$ unless the initial guess is close to the solution. Our implementation uses the novel W4 method \citep{Okawa+22} that shows better global convergence than Newton-Raphson-based methods.

To exclude a source of recombination energy, we remove its contribution from equation (\ref{eq:epsilon}). In total, we compare simulations with five different implementations of recombination:
\begin{enumerate}
	\item \irecA: We include the recombination energy of helium and hydrogen, including the energy released during \molH molecular formation. All terms in equation (\ref{eq:epsilon}) are retained.
	\item \irecB: We exclude the energy contributed by H${}_2$ formation, including only the last three terms in equation (\ref{eq:epsilon}).
	\item \irecC: We only include helium recombination energy (from both the singly- and doubly-ionised states), retaining only the last two terms in equation (\ref{eq:epsilon}).
	\item \irecD: We do not allow any recombination energy to be deposited into the gas ($\varepsilon_\mathrm{ion} = 0$), but include the increase/decrease in mean molecular weight associated with recombination/ionisation by retaining the dependence of $\mu = \mu(x_i)$ on the ionisation fractions $x_i$.
	\item \irecE: We neglect all effects of recombination, setting $\varepsilon_\mathrm{ion}=0$ and fixing the mean molecular weight at $\mu = 0.62$, which corresponds to a fully ionised envelope with the same composition. This is identical to the ``gas + radiation'' \ac{EoS} simulation in Paper I \citep{Lau+22}.
\end{enumerate}

\begin{table}
	\centering
	\begin{tabular}{llll}
		\toprule
		(i) Model & (ii) $a_\mathrm{f}$ /\Rsun & (iii) $f_{\rm{k}+\rm{p}+\rm{th}}$ & (iv) $f_{\rm{k}+\rm{p}}$ \\ \midrule
		 \irecA   & 44.8 (37.7) & 0.62 (0.77) & 0.43 (0.64) \\
		 \irecB   & 44.7 (39.3) & 0.56 (0.65) & 0.37 (0.54) \\
		 \irecC   & 44.7 (40.0) & 0.40 (0.48) & 0.31 (0.45) \\
		 \irecD   & 38.6 (34.6) & 0.30 (0.36) & 0.25 (0.32) \\
		 \irecE   & 37.2 (34.0) & 0.29 (0.40) & 0.25 (0.37) \\
		\bottomrule
	\end{tabular}
	\caption{Summary of our simulation results, with those obtained at lower resolution in parentheses. From left to right, the columns list (i) The different recombination energy sources included in the simulation, as described in Section \ref{subsec:recombination}, (ii) The final separation, defined at a reference time when the ratio of the orbital period to the inspiral time-scale falls below a threshold ($-P_\text{orb}\dot{a}/a < 5\times10^{-4}$). Columns (iii) and (iv) report the fraction of unbound envelope mass at that same reference point, with $f_{\rm{k}+\rm{p}}$ assuming a purely mechanical criterion for considering material to be unbound and $f_{\rm{k}+\rm{p}+\rm{th}}$ also including thermal energy (see Section \ref{sec:results}).}
	\label{tab:summary}
\end{table}

\subsection{The amount of available recombination energy}
\begin{figure}
    \includegraphics[width=\linewidth]{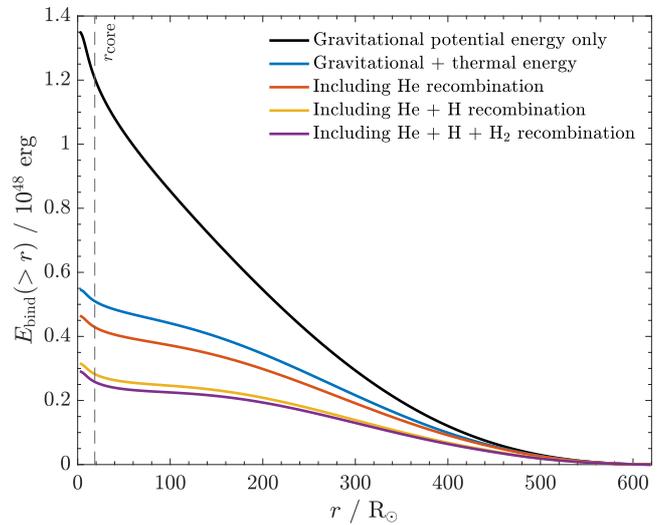}
    \caption{Comparison of the cumulative binding energy integrated from the surface (equation (\ref{eq:ebind})) when different sources of energy are included. The black line includes the gravitational potential energy only, while the blue line also includes thermal energy (first two terms of equation (\ref{eq:eint})). The remaining lines successively include the following sources of recombination energy: helium recombination energy (red line), hydrogen atomic recombination energy (yellow line), and hydrogen molecular recombination energy (purple line). The vertical dashed line marks the location of the core boundary in our simulations.}
    \label{fig:ebind}
\end{figure}

To obtain an upper limit to the role of recombination energy and understand the relative importance of different recombination energy sources, we plot in Figure \ref{fig:ebind} the donor's binding energy profile when including different sources of recombination energy. We integrate the binding energy inwards from the surface as given by
\begin{align}
	E_\text{bind}(>r) = \int_{M}^{m(r)} \bigg[
	\varepsilon_\mathrm{th}(m') + \varepsilon_\mathrm{ion}(m') - \frac{Gm'}{r'}
	\bigg]dm',
	\label{eq:ebind}
\end{align}
where $\varepsilon_\mathrm{th}$ is the thermal energy, $\varepsilon_\mathrm{ion}$ is the ionisation potential from equation (\ref{eq:epsilon}), and the third term in the integrand is the gravitational potential. The sum $u = \varepsilon_\mathrm{th} + \varepsilon_\mathrm{ion}$ is the specific internal energy, equivalent to equation (\ref{eq:eint}). The vertical dashed line marks the location of our simulations' core boundary, $r_\mathrm{core} = 18.5\Rsun$, which represents the base of the convective envelope. The value of $E_\mathrm{bind}(>r)$ at $r=r_\mathrm{core}$ is therefore the convective envelope's binding energy. The total gravitational binding energy is $1.2\times10^{48}\erg$. If all of the envelope's thermal energy contributes to ejecting the envelope, the binding energy is reduced by more than a factor of two, giving $5.1\times10^{47}\erg$ (the thermal energy for a gas-pressure dominated envelope is half the magnitude of the gravitational potential energy, and approaches the magnitude of the gravitational potential energy as radiation pressure becomes dominant).

The total ionisation potential energy in the envelope is $2.5\times10^{47}\erg$, of which $1.5\times10^{47}\erg$ (58 per cent) is from ionised hydrogen and $8.1\times10^{46}\erg$ (32 per cent) is from ionised helium. The contribution from the potential of \molH formation, $2.4\times10^{46}\erg$, is comparatively small (9.7 per cent). Applying the energy formalism, if recombination may be fully and efficiently used to eject the envelope, including helium recombination energy can increase the final separation by 16 per cent, while including both hydrogen and helium recombination can increase the final separation by 45 per cent, compared to if only thermal energy were fully used.

However, we note a number of ways by which the expectations outlined above are simplified. Firstly, unlike thermal energy, recombination energy is released at specific temperatures: $\approx$ 6,000 K for hydrogen recombination, 13,000 K for HeII recombination, 40,000 K for HeIII recombination, and 1,300 K for \molH molecular formation. If the initial adiabatic expansion driven by orbital energy deposition does not sufficiently cool the envelope to a recombination temperature, the associated recombination energy will not be released. Even if recombination energy were released in the envelope, its role in ejecting the envelope depends on where and when this energy is injected. For example, recombination may lead to runaway ejection if the binding energy at the partial ionisation zone is less than the recombination energy released there \citep{Ivanova+15,Ivanova&Nandez16}. Recombination may also induce unstable pulsations during the self-regulated phase that could eject the whole envelope \citep[][although we only model the dynamical phase]{Clayton:2017}. Finally, the contribution of recombination energy to the envelope binding energy depends on the envelope bifurcation point, diminishing with a deeper boundary where the magnitude of the gravitational potential may greatly exceed the ionisation potential \citep{Kruckow+16}.


\section{Results}
\label{sec:results}
We examine two key quantities that characterise the outcome of a \ac{CE} simulation: the fraction of unbound envelope mass (Section \ref{subsec:unbound}) and the final separation (Section \ref{subsec:sep}). We consider a gas parcel as unbound if it has positive energy, and either use a ``mechanical'' criterion that sums the specific kinetic and gravitational potential energies ($e_\mathrm{k} + e_\mathrm{p}>0$) or a ``thermal'' criterion that also includes thermal energy ($e_\mathrm{k} + e_\mathrm{p} + e_\mathrm{th} > 0$), excluding recombination energy. Table \ref{tab:summary} lists each simulation's final separation, $a_\mathrm{f}$, and fraction of unbound envelope mass calculated with the thermal ($f_\mathrm{k+p+th}$) and mechanical criteria ($f_\mathrm{k+p}$), defined at a reference time where the inspiral time-scale is much longer than the orbital period, $-P_\mathrm{orb}\dot{a}/a < 5\times10^{-4}$, where $a$ is the semi-major axis and $P_\mathrm{orb}$ is the orbital period.

We primarily reference $f_\mathrm{k+p+th}$ in the ensuing discussion. Although this thermal criterion assumes that thermal energy is eventually fully converted into mechanical work, which is not guaranteed, it yields an amount of unbound envelope mass that is closer to its asymptotic value \citep[see our discussion in Paper I,][]{Lau+22}. We plot the evolution of the fraction of unbound envelope mass and the core-companion separation as functions of a shifted time, $t-t_0$, since the companion reaches 75 per cent of the initial donor radius (Figures \ref{fig:unbound} and \ref{fig:sep}).

\subsection{Unbound envelope mass}
\label{subsec:unbound}

\begin{figure*}
    \includegraphics[width=0.9\linewidth]{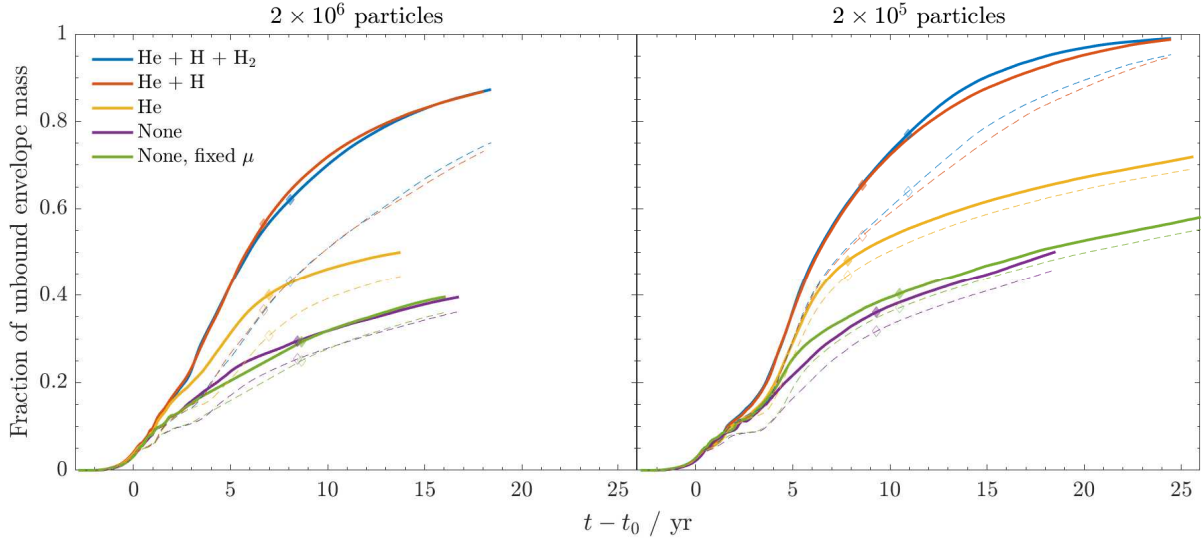}
    \caption{Comparison of the fraction of unbound envelope mass for simulations including different sources of recombination energy, as explained in Section \ref{subsec:recombination}. \textit{Left panel:} Results from our default simulations with $2\times 10^6$ \ac{SPH} particles. \textit{Right panel:} Results from simulations that use ten times fewer \ac{SPH} particles ($2\times 10^5$ particles), but are otherwise identical. The horizontal axis plots the time, $t-t_0$, since the companion plunges beneath 75 per cent of the initial donor radius. The diamond markers indicate where $-P_\mathrm{orb}\dot{a}/a$ falls beneath $5\times10^{-4}$, which is also the reference time we use to define a post-plunge-in semi-major axis, $a_\mathrm{f}$ (see Section \ref{subsec:sep}). The solid lines consider a gas parcel as unbound if the sum of its kinetic, potential, and thermal energy is positive ($e_\mathrm{k} + e_\mathrm{p} + e_\mathrm{th} > 0$), while the dashed lines exclude thermal energy from the criterion (only requiring $e_\mathrm{k} + e_\mathrm{p} > 0$).}
    \label{fig:unbound}
\end{figure*}

Figure \ref{fig:unbound} shows the evolution of the fraction of unbound envelope mass for the simulations listed in Section \ref{subsec:recombination}. For all simulations, material is still gradually becoming unbound at a rate of $\approx 0.1$\Msun / yr more than a decade after the dynamical plunge. Due to their large computational expense, we were unable to run the simulations long enough to find the final amount of unbound mass. Although, the simulations at lower resolution (using $2\times10^5$ \ac{SPH} particles) extend to $t-t_0 = 25$ yr. To compare the simulations, we therefore use the values of $f_\mathrm{k+p+th}$ in Table \ref{tab:summary} as references, which are indicated with filled diamond markers in Figure \ref{fig:unbound}.

Excluding recombination energy (\irecD and \irecE) results in the smallest amount of ejected material ($f_\mathrm{k+p+th}=0.30$ and $f_\mathrm{k+p+th}=0.29$, respectively). On the other hand, both the \irecA and \irecB simulations eject at least 87 per cent of the envelope by the end ($f_\mathrm{k+p+th}=0.62$ and 0.56, respectively), with the low-resolution results suggesting complete envelope ejection may be possible after another decade.

The simulation including only helium recombination energy ejects at least half of the envelope mass by the end, with $f_\mathrm{k+p+th}=0.40$, which is 33 per cent larger than $f_\mathrm{k+p+th}$ for the simulation without recombination energy (\irecD). On the other hand, comparing the \irecB and \irecC cases reveals that adding hydrogen recombination energy increases $f_\mathrm{k+p+th}$ by another 40 per cent, compared to the case where only helium recombination energy is included. This additional amount of unbound ejecta sets an upper bound on the role of hydrogen recombination for the simulated system, as our adiabatic simulations assume all recombination energy is locally thermalised. In reality, a significant fraction of hydrogen recombination energy may instead be radiated away, although we suggest in Section \ref{subsec:transport} that this may mainly occurs in ejecta that are already unbound.

The fractions of unbound mass in the \irecA and \irecB simulations are very similar at any given time, implying \molH formation plays an insignificant role in \ac{CE} ejection. Their corresponding curves in Figure \ref{fig:unbound} are almost identical up to $t-t_0=6$ years after the plunge-in, because less than 1 per cent of the envelope mass in each case has cooled to temperatures conducive to \molH recombination ($\approx 1,300$ K). But by the time ejecta can be adiabatically cooled to these temperatures, we find that they have already substantially expanded and become unbound. Even if the recombination energy of \molH is injected into an envelope that is still bound, it constitutes less than 5 per cent of the envelope binding energy (see Section \ref{subsec:recombination} and Figure \ref{fig:ebind}). Similarly, the role of dust-driven acceleration for \ac{CE} ejection is expected to be insignificant in the system and time-scales we simulate. An upper limit may be obtained by assuming the initial stellar luminosity, $L\sim 10^{38}~\mathrm{erg}~\mathrm{s}^{-1}$, is fully used to accelerate the entire envelope ($M_\mathrm{env}\sim 10^{34}~\mathrm{g}$) over a duration $\Delta t \sim 100\yr$, taken to be much longer than the \ac{CE} phase we simulate. The envelope velocity increases by $L\Delta t / (M_\mathrm{env}c) \sim 10^{-2}\kms$, a negligible amount compared to the donor's surface escape velocity, $\approx 90\kms$ \citep[see also][who focus on a $1\Msun$ giant branch star and require $\Delta t = 1.3\times 10^5\yr$ to eject the envelope]{Glanz&Perets18}. While unimportant for envelope ejection, molecule and even dust formation may influence the formation of the surrounding nebula accelerated by winds from the stripped donor star in the post-\ac{CE} binary \citep[e.g.][]{Pejcha+16}.

Comparing the \irecD and \irecE simulations allows us to isolate the effect of increased mean molecular weight, $\mu$, associated with recombination. Figure \ref{fig:unbound} shows that there are no significant differences in the fraction of unbound envelope mass, and so recombination mainly impacts the amount of unbound material via its energetic contribution.

\begin{figure*}
    \includegraphics[width=0.9\linewidth]{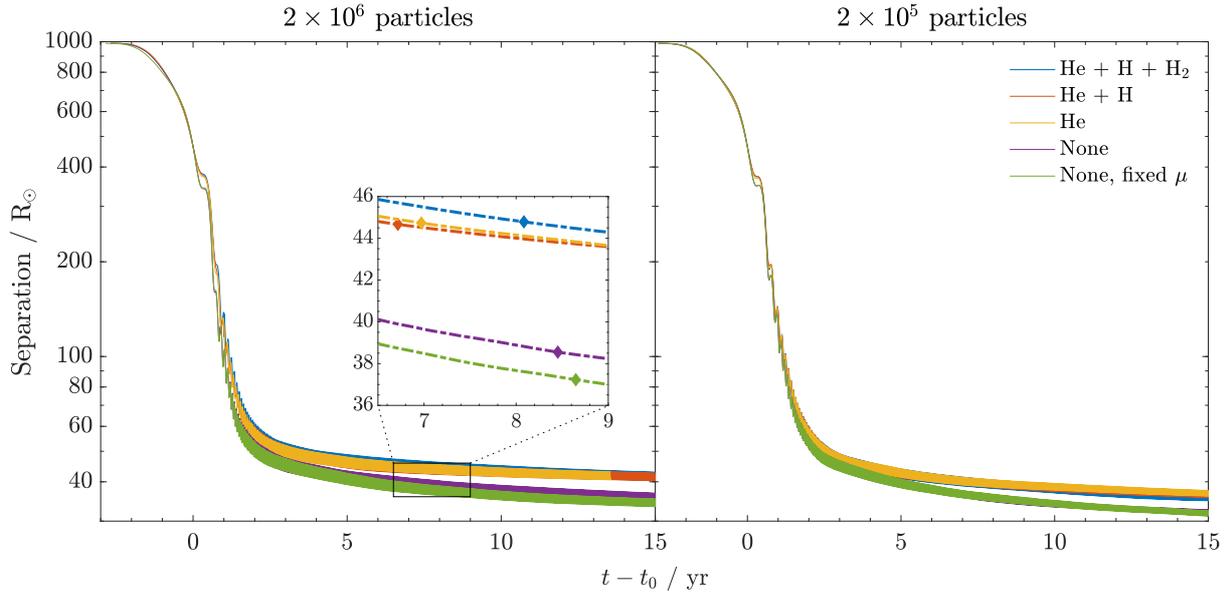}
    \caption{Same as Figure \ref{fig:unbound}, but instead showing the evolution in core-companion separation. In the left panel, we show a magnified region between $t-t_0=$ 6.5 and 9.0 yr to highlight model differences. The magnified region shows the orbital semi-major axis, $a$, of the stellar cores instead of the instantaneous separation. The diamond markers indicate the orbital semi-major axis, $a_\mathrm{f}$, at which $-P_\mathrm{orb}\dot{a}/a$ falls beneath $5\times10^{-4}$.}
    \label{fig:sep}
\end{figure*}

Figure \ref{fig:unbound} also reveals finite-resolution effects through comparison with simulations performed with ten times fewer \ac{SPH} particles, shown as the faded lines. At lower resolution, less material is ejected near the dynamical plunge-in ($t-t_0\approx 0$), but more material is ejected during the slow spiral-in ($t-t_0 \gtrsim 3$). The first effect can be understood as follows. During the dynamical plunge-in, the energy injection mechanism is shock heating, where material becomes unbound after it is swept by the companion's bow shock one or more times. The shock front is more finely resolved and therefore heats gas in a more concentrated region with higher resolution, leading to a higher fraction of unbound \ac{SPH} particles. The second effect is because the companion spirals in deeper during the plunge-in at lower resolutions (see Section \ref{subsec:sep} and Figure \ref{fig:sep}). This results in relatively more energy injection (steeper increase in the fraction of unbound envelope mass), seen near $3 \lesssim (t-t_0)/\yr \lesssim 6$ in Figure \ref{fig:unbound}.

The fraction of unbound envelope mass calculated with the purely mechanical ($e_\mathrm{k}+e_\mathrm{p}>0$) criterion (dashed lines) is $\approx5-20$ per cent smaller than that calculated with our default criterion including thermal energy ($e_\mathrm{k}+e_\mathrm{p}+e_\mathrm{th}>0$) (solid lines), which is similar to \cite{Moreno+22}. We also observe that the discrepancy in the amount of unbound mass calculated with the two criteria is greater at higher resolution, indicating that the envelope's thermal to kinetic energy ratio is resolution-dependent. This may be because that at higher resolution, more energy is stored as thermal energy in the convective eddies during the slow spiral-in (see Section \ref{subsec:ejecta}). At lower resolutions, where these flows are less well resolved, orbital energy is more readily converted into kinetic energy.

\subsection{Final separation}
\label{subsec:sep}
Figure \ref{fig:sep} shows that the rates of orbital shrinkage during the dynamical plunge are similar across different simulations, shrinking the orbit to $\approx40-60\Rsun$ within 3 years. In our simulations, the orbital semi-major axis of the stellar cores continue to decrease gradually by a few times $0.1\Rsun~\mathrm{yr}^{-1}$ even nearly a decade after the dynamical plunge-in. We therefore define the final separation, $a_\mathrm{f}$, at a reference time where the inspiral time-scale becomes 2,000 times longer than the orbital period, $-P_\mathrm{orb}\dot{a}/a < 5\times10^{-4}$. The final separations, $a_\mathrm{f}$, of all our simulations are listed in Table \ref{tab:summary}.

These separations, along with the magnified region shown in Figure \ref{fig:sep}, show that the final separations for the five different models cluster around two distinct values. The simulations that include at least one source of recombination energy (\irecA, \irecB, and \irecC) have similar final separations ($a_\mathrm{f} = 44.7-44.8\Rsun$) that are approximately 16 per cent higher than models that do not (\irecD and \irecE, with $a_\mathrm{f} = 38.6$ and 37.2\Rsun, respectively).

The larger final separations in \ac{CE} simulations performed with recombination energy is consistent with a number of previous studies \citep{Sand+20,Lau+22,Gonzalez-Bolivar+22}, supporting the idea that the additional envelope expansion driven by recombination energy helps stall the dynamical plunge-in earlier. However, these studies did not distinguish the effects of different recombination energy sources. Our results suggest helium recombination rather than hydrogen recombination as the main cause of the larger final separation. This can be inferred from the fact that the addition of hydrogen recombination energy in the \irecB simulation, despite ejecting 40 per cent more material relative to the \irecC simulation, does not halt the dynamical plunge-in earlier. The \irecB and \irecC simulations have similar final separations, $a_\mathrm{f}=44.7\Rsun$ in both cases. This could be because most of the hydrogen recombination energy is injected after the plunge-in and in the outer, marginally-bound parts of the \ac{CE} that are energetically decoupled from the inner binary \citep{Nandez&Ivanova16}. Helium recombination, however, occurs much deeper in the \ac{CE}, where most of the orbital energy is released. On the other hand, \cite{Ivanova&Nandez16} and \cite{Reichardt+20} do not find that recombination energy changes the final separation significantly in their simulations. These varying findings in the literature are conceivably due to differences in starting models. For example, donor stars with different masses and at different evolutionary stages have partial ionisation zones located at different relative mass coordinates.

The \irecA and \irecB simulations have similar final separations ($a_\mathrm{f} = 44.8$ and 44.7\Rsun, respectively), consistent with our finding in Section \ref{subsec:unbound} that \molH formation plays an insignificant role.

The simulations conducted with ten times fewer \ac{SPH} particles have final separations that are around 10 per cent smaller (Figure \ref{fig:sep}, right panel). However, the relative differences between each simulation are consistent with our discussion above for the default simulations. Particularly, the finding that helium recombination energy increases $a_\mathrm{f}$, as opposed to hydrogen recombination energy, still holds.

\section{Discussion} \label{sec:discussion}
\subsection{Impact of recombination on the ejecta}
\label{subsec:ejecta}
\begin{figure*}
    \includegraphics[width=.94\linewidth]{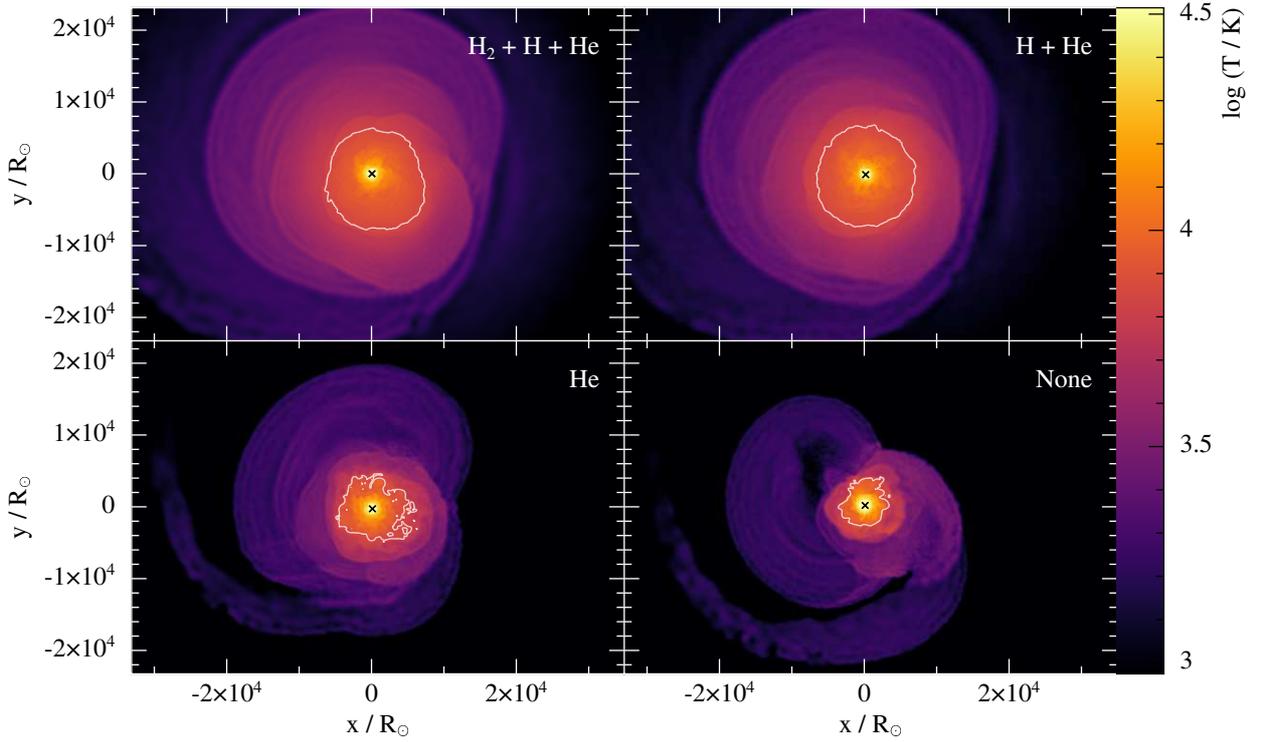}
    \caption{Comparison of the ejecta temperature in the $z=0$ (face-on) cross-section across the various simulations at $t=9.97$ yr, showing more expanded and uniform-temperature ejecta where hydrogen recombination has been included (simulations in the top panels). The white line is a contour of constant hydrogen ionisation fraction, $1-[\mathrm{HI}] = 0.7$, indicating where hydrogen recombination is actively occurring. The black crosses indicate the locations of the stellar cores. All simulation renderings in this paper were created with \SPLASH \citep{Price07}. Videos of our simulations are available at \url{https://themikelau.github.io/CE_recombination}.}
    \label{fig:T_xy}
\end{figure*}

\begin{figure*}
    \includegraphics[width=.94\linewidth]{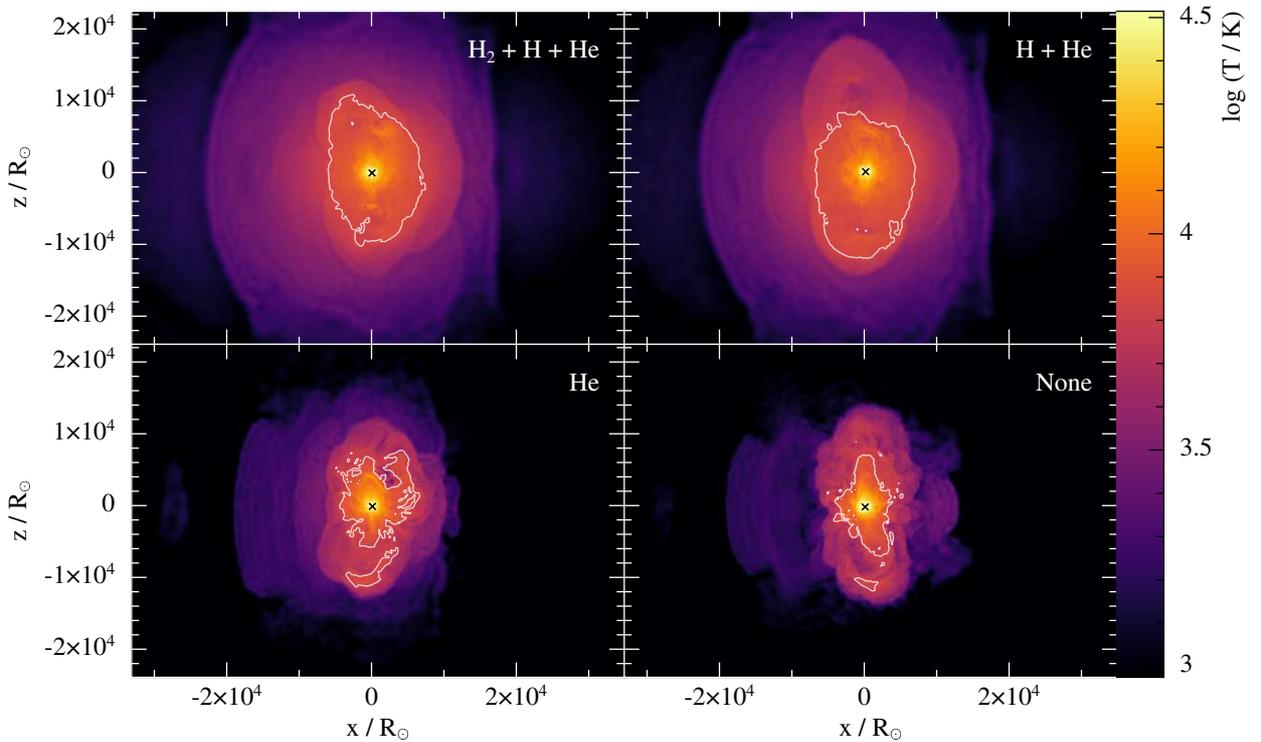}
    \caption{Same as Figure \ref{fig:T_xy}, but showing the $y=0$ (edge-on) cross-section.}
    \label{fig:T_xz}
\end{figure*}

We compare the impact of different types of recombination on the ejecta structure and morphology. Figures \ref{fig:T_xy} and \ref{fig:T_xz} show temperature slices of the \ac{CE} across different simulations at $t=9.97$ yr (around $t-t_0 = 7.3$ yr, depending on the simulation). The ejecta form an asymmetric structure consisting of hot, low-density bipolar cocoons expanding at $\approx 10-15\kms$. As seen from the edge-on slices shown in Figure \ref{fig:T_xz}, these lobes are around $10^4\Rsun$ in extent, and enshrouded in unbound material that was ejected prior to the plunge-in \citep[see also][]{MacLeod+18,Ondratschek+22}. These cocoons are driven by 100--120\kms outflows launched near the stellar cores, perpendicular to the orbital plane. These outflows are hydrodynamically collimated by comparatively dense and hot, bound material in the equatorial region that are about $\sim 10^3\Rsun$ in radius. Convective mixing develops in this bound material, as shown in the orbital plane slices in Figure \ref{fig:T_xy}. The formation of bipolar outflows is qualitatively consistent with observations of post-\ac{CE} nebulae \citep[e.g.][]{Kaminski+18}, but a proper comparison would require modelling the expansion of the nebula driven by irradiation and stellar winds from the stripped core \citep{Garcia-Segua+18,Frank+18,Zou+20}.  

In Figures \ref{fig:T_xy} and \ref{fig:T_xz}, there are visible differences between the upper and the lower panels. The simulations in the upper panels include hydrogen recombination energy, which causes the coolest material to expand past $\approx 2\times 10^4\Rsun$ from the centre, and yet also retain higher temperatures (though the temperatures remain unrealistically high in our simulations due to the absence of radiative cooling, which should be particularly efficient after hydrogen recombination). The white line is a contour of constant hydrogen ionisation fraction, $1-[\mathrm{HI}] = 0.7$, which we show in Section \ref{subsec:transport} encloses the region where most useful recombination energy is injected. This contour is more spherical and uniform in temperature (at the recombination temperature $\log(T~/~\mathrm{K}) \approx 3.8$), and encloses a larger region than in the lower panels. We ascertain that this is a result of hydrogen recombination energy, rather than orbital energy injected by the stellar cores, from observing that the temperature slices of the \irecC simulation look qualitatively different from those of the \irecB simulation, despite having nearly the same final separations.

The \irecC and \irecD simulations are similar to each other, both displaying stronger convective mixing, leading to greater temperature inhomogeneity in the ejecta. Convection in the bound ejecta is induced by gradual heating by the slowly in-spiralling stellar cores after the dynamical phase \citep{Lau+22,Moreno+22}. The greater amount of bound and infalling material in the \irecC and \irecD cases could contribute to the more prominent mixing. There also appears to be slightly more convective plumes in the \irecD case, which could be due to the $\approx16$ per cent smaller final separation of the stellar cores, which must have injected more energy to the surrounding medium.

\subsection{The ability for recombination energy to thermalise}
\label{subsec:transport}

\begin{figure}
    \includegraphics[width=1.02\linewidth]{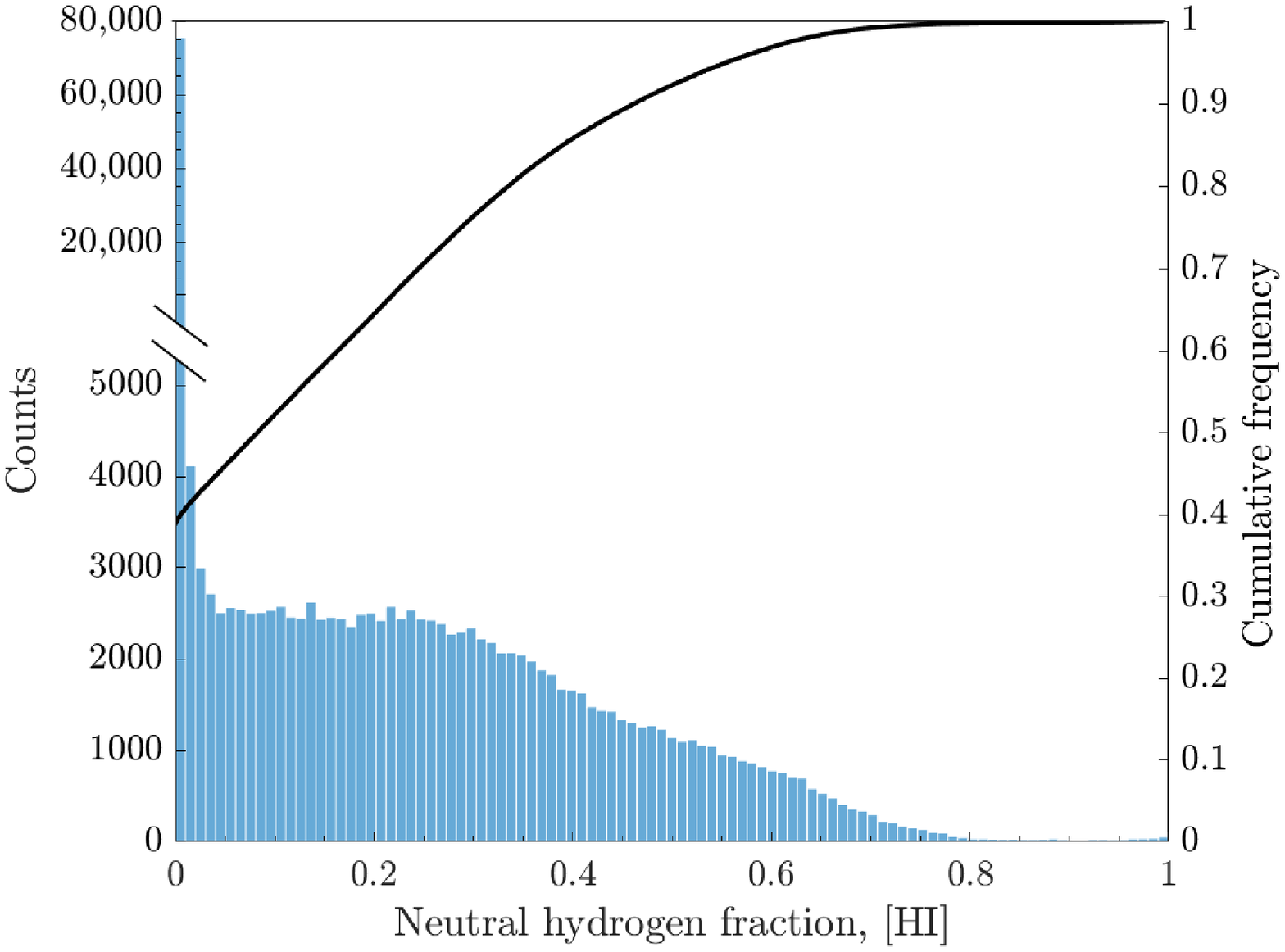}
    \caption{Histogram of the neutral hydrogen fraction, [HI], of \ac{SPH} particles at the moment they are marked as unbound (attaining positive energies). The black line shows the cumulative frequency. We break the left vertical axis between 6,000 and 10,000 counts to accommodate for the large amount of material ejected while hydrogen is still fully ionised.}
    \label{fig:hi_fraction}
\end{figure}

Thermal energy injected by the plunging companion's bow shock or by recombination can either perform work to expand the envelope or be transported away by processes such as radiation and convection, in the absence of other heat sources and sinks. Our adiabatic simulations might overestimate the amount of work done on the envelope by these processes as there is no radiation transport. A number of existing works estimate the ability for recombination energy to expand the \ac{CE} \citep{Sabach+17,Grichener+18,Ivanova18,Soker+18}, which is a matter of ongoing discussion. \cite{Wilson&Nordhaus19,Wilson&Nordhaus20} have also suggested that the post-\ac{CE} separation could be determined by the boundary at which convection is able to transport away the luminosity injected into the envelope during the dynamical plunge-in. More recently, their extension to \acp{CE} with massive stellar donors shows that the envelope expansion time-scale is short compared to the convective transport time-scale computed with static 1D stellar models \citep{Wilson&Nordhaus22}.

In our simulations, helium recombination occurs on the time-scale of the plunge-in, as the plunge-in directly drives the expansion of the envelope through the helium partial ionisation zone. So if convective transport is unlikely to take away orbital energy, it is also unlikely to take away helium recombination energy. Helium recombination energy is also released in deep, optically-thick layers of the \ac{CE}, and so cannot be transported away by radiative diffusion. Our \irecC simulation, which finds a 16 per cent larger final separation and 33 per cent more ejected material, therefore likely represents a lower limit to the efficacy of recombination energy to help eject a \ac{CE} for our chosen binary system.

On the other hand, the existing literature suggests hydrogen recombination energy to be less efficiently used, as a substantial portion is injected in optically-thin layers, where it can be radiated away \citep{Grichener+18}. Yet, this does not directly imply that our adiabatic simulations overestimate the amount of ejected material, if the recombination energy that should have been radiated away is released in material that is already unbound. To see if this is true, we plot in Figure \ref{fig:hi_fraction} the distribution of neutral hydrogen fraction, [HI], recorded at the moment when each \ac{SPH} particle is marked as unbound according to our criterion\footnote{It is possible for a particle to transition between states of negative and positive total energy more than once. In that case, we record the hydrogen ionisation fraction during the particle's final transition from negative to positive total energy.}. We display results for the \irecB simulation conducted with $2\times10^5$ \ac{SPH} particles, which completely ejects the envelope by the end of runtime. The counts across all bins therefore sum up to $2\times10^5$. The black line shows the normalised cumulative count. The distribution is dominated by a peak at [HI] $=0$ (containing 38 per cent of the envelope mass), which includes all material that was ejected without any use of hydrogen recombination energy. The distribution is flat up to [HI] $\approx 0.3$ (containing 80 per cent of the envelope mass), beyond which it linearly declines to [HI] $\approx 0.8$. This shows that a relatively small fraction of the available hydrogen recombination energy (median of 0.085) produces the 40 per cent more unbound material seen in Figure \ref{fig:unbound}.

Previous works studying thermally expanding, 1D hydrostatic stellar models have shown that these largely ionised layers ([HI] $\lesssim 0.5$) are sufficiently optically-thick that photon diffusion is unlikely to prevent recombination energy from driving expansion \citep{Grichener+18,Soker+18}. The energy released at lower ionisation fractions is not responsible for the additional amount of ejected material in the \irecB simulations, and would instead contribute to the H$\alpha$ emission of luminous red novae associated with \acp{CE} \citep{Ivanova+13b,MacLeod+17a,Matsumoto&Metzger22}. \cite{Grichener+18} instead suggest that convective transport might be efficient in transporting away hydrogen recombination energy in more ionised layers. However, this hydrogen partial ionisation zone is marginally unbound, outflowing at significant fractions of the sound speed, and significantly out of hydrostatic equilibrium, and so departs from the regime of applicability of mixing length theory. Convective transport is therefore expected to be much less efficient than in a 1D, hydrostatic profile. It would be valuable to self-consistently model convection in a \ac{CE} by incorporating radiation transport into 3D simulations \citep[see, e.g.,][]{Ricker+2018}

\section{Summary and conclusions} \label{sec:conclusion}
We studied the role of different sources of recombination energy on the \ac{CE} evolution experienced by a massive star donor. We performed a series of 3D hydrodynamical simulations involving a 12\Msun \ac{RSG} donor with a 3\Msun companion, using the setup from our previous work \citep{Lau+22}. These simulations include different sources of recombination energy, assumed to thermalise locally in the envelope. By comparing these simulations, we have been able to infer the distinct effects of hydrogen and helium recombination. We list our main findings below:

\begin{enumerate}
	\item Helium recombination energy, comprising 32 per cent of the envelope's total recombination energy, leads to ejecting 33 per cent more envelope mass, compared to when recombination energy is not included at all. This energy is released deep in the envelope, where it is unlikely to be transported away by radiative diffusion or convection, and so represents a likely lower limit to the use of recombination energy in \ac{CE} evolution.
	\item Our simulations including both hydrogen and helium recombination eject at least 87 per cent of the envelope upon termination. Mass is still becoming unbound at a rate of $\approx 0.1$\Msun / yr. Simulations including hydrogen and helium recombination energy eject the full envelope when evolved at lower resolution for another decade. Hydrogen recombination therefore leads to 40 per cent more ejected envelope mass compared to just including helium recombination energy.
    \item Only a small fraction of hydrogen recombination energy is required to eject this additional unbound material. The majority (80 per cent) of the ejecta becomes unbound when only less than 30 per cent of the hydrogen recombination energy has been released. In these layers, which are more ionised and therefore more optically-thick, a large fraction of recombination radiation is expected to thermalise in the envelope and radiative diffusion is inefficient.
	\item Molecular recombination into \molH plays an insignificant role in \ac{CE} ejection, because almost all the material that is able to adiabatically cool to $\approx$ 1,300 K is already unbound. The energy available from molecular recombination is also limited, comprising only 5 per cent of the envelope binding energy for our \ac{RSG} donor.
	\item Helium recombination increases the final separation by $\approx 16$ per cent. This is because helium recombination occurs deeper in the \ac{CE} where most of the orbital energy is released, unlike hydrogen recombination. Adding hydrogen recombination does not significantly alter the final separation.
	\item The ejecta in our simulations contain hot, low-density cocoons inflated by bipolar outflows. The deposition of hydrogen recombination energy results in more extended and spherically-symmetric ejecta, and reduces the amount of convective mixing in the bound material near the stellar cores. 
\end{enumerate}

\section*{Acknowledgements}
We thank Orsola De Marco and Miguel Gonz\'{a}lez-Bol\'{i}var for useful discussions. M. Y. M. L. acknowledges support by an Australian Government Research Training Program (RTP) Scholarship. IM is a recipient of the Australian Research Council Future Fellowship FT190100574. Parts of this research were supported by the Australian Research Council Centre of Excellence for Gravitational Wave Discovery (OzGrav), through project number CE170100004, and in part by the National Science Foundation under Grant No. NSF PHY-1748958. Parts of the simulations presented in this work were performed on the Rusty supercomputer and Popeye supercomputer of the Flatiron Institute, which is supported by Simons Foundation, and on the Gadi supercomputer of the National Computational Infrastructure (NCI), which is supported by the Australian Government, and on the OzSTAR national facility at the Swinburne University of Technology. The OzSTAR program receives funding in part from the Astronomy National Collaborative Research Infrastructure Strategy (NCRIS) allocation provided by the Australian Government.

\section*{Data availability}
The data used to produce all figures in this article are available on Monash University's Bridges repository, at \url{https://dx.doi.org/10.26180/20418837.v2}.






\bibliographystyle{mnras}
\bibliography{bibliography.bib}

\appendix



\bsp	
\label{lastpage}
\end{document}